\documentclass[aps,prl,superscriptaddress,twocolumn]{revtex4-1}
\usepackage[pdftex]{graphicx}
\usepackage{caption}
\usepackage{subfigure}

\begin{document}

%\preprint{}

\title{Replenish and relax: explaining logarithmic annealing in disordered materials}


\author{Laurent Karim B\'{e}land}
\email[]{laurent.karim.beland@umontreal.ca}
\author{Yonathan Anahory}
\email[]{y.anahory@gmail.com}
\author{Dries Smeets}
%\email[]{dries.smeets@gmail.com}
\author{Matthieu Guihard}
\author{Peter Brommer}
\email{p.brommer@warwick.ac.uk}
\altaffiliation[Current address: ]{Department of Physics and Centre for Scientific Computing, University of Warwick, Gibbet Hill Road, Coventry CV4 7AL, United Kingdom}
\author{Jean-Fran\c{c}ois Joly}
\email[]{jeanf.joly@gmail.com}
\author{Jean-Christophe Pothier}
\email[]{jc.pothier@phytronix.com}
\author{Laurent J. Lewis}
\email[]{laurent.lewis@umontreal.ca}
\author{Normand Mousseau}
\email[]{normand.mousseau@umontreal.ca}
\author{Fran\c{c}ois Schiettekatte}
\email[]{francois.schiettekatte@umontreal.ca}

%\email[]{jeanf.joly@gmail.com}

%\email[]{jc.pothier@phytronix.com}

%

%\thanks{}
\affiliation{Regroupement Qu\'{e}b\'{e}cois sur les Mat\'{e}riaux de Pointe(RQMP), \\
D\'{e}partement de physique, Universit\'{e} de Montr\'{e}al, \\
Case Postale 6128, Succursale Centre-ville, Montr\'{e}al, Qu\'{e}bec, H3C 3J7,Canada}


\date{\today}

\begin{abstract}

Fatigue and aging of materials are, in large part, determined by the evolution of the atomic-scale structure in response to strains and perturbations. This coupling between microscopic structure and long time scales remains one of the main challenges in materials study. Focusing on a model system, ion-damaged crystalline silicon, we combine nanocalorimetric experiments with an off-lattice kinetic Monte Carlo simulation to identify the atomistic mechanisms responsible for the structural relaxation over long time scales. We relate the logarithmic relaxation, observed in a number of systems, with heat-release measurements. The microscopic mechanism associated with logarithmic relaxation can be described as a two-step replenish and relax process. As the system relaxes, it reaches deeper energy states with logarithmically growing barriers that need to be unlocked to replenish the heat-releasing events leading to lower energy configurations.

\end{abstract}

%\pacs{1}
%\keywords{}

\maketitle

%\section{}
%\section{Introduction}
%\subsubsection{}

The importance of defects in determining material properties and evolution can hardly be
overstated, and yet understanding their evolution over experimental timescales remains a
challenge. This problem affects materials ranging from somewhat simple crystalline systems
to glasses, alloys and cements and is at the roots of phenomena affecting materials
reliability, aging and fatigue. Over the years, their evolution has been characterized by
various models based on continuous, mesoscopic or atomic scale models and generally fitted
to experimental data~\cite{Andreozzi2003, Chen2007}. Direct comparison between models,
experiments, and atomic-scale simulations, however, have been limited because of the often
extensive time scale difference between the three, leaving considerable space for
interpretation in the fitting procedure even when describing apparently simple relaxation
processes such as logarithmic relaxation observed in polymer glasses, for
example~\cite{McKenna2003, Knoll2009}, and in ion-implanted amorphous and crystalline
systems~\cite{1aN9,Kallel10}. Here, we select ion implantation induced disorder in
monocrystalline silicon (\emph{c}-Si), a controlled approach for perturbing samples that
can be closely reproduced numerically ~\cite{Caturla96,Beck08}. Combined with calorimetric
measurements, ion implantation is an ideal technique for studying the kinetics and
thermodynamics of complex structures evolution in a systematic way.

Many models have been proposed for describing the accumulation and annealing of
ion-induced disorder in materials. The Frenkel pair model, for example, describes the
disorder in terms of isolated vacancies (V) and interstitials (I) \cite{SRIM}; in
contrast, MD simulations show that the majority of defects induced by the impact of a
single ion of a few keV are found in the form of heavily-damaged zones commonly called
amorphous pockets (APs)~\cite{Caturla96,Nordlund,Pothier:2011p2} that anneal in steps,
with a wide range of activation energies that depend on the details of each APs' interface
with the crystal \cite{Caturla96}, as observed experimentally \cite{Donnelly03}. A popular
model to account for damage accumulation proposed by Marqu\'{e}s and coworkers
\cite{Marques03} describes APs as an assembly of \emph{bound defects} corresponding to
IV-pairs~\cite{2aN7}. While never detected directly in experiments, this mechanism was
observed in various forms in short-time simulations~\cite{2aN8,2aN9,Kallel10}. As shown
below, this model does not describe nanocalorimetriy (NC) measurements correctly and we
are still seeking the correct atomic-scale relaxation model for implanted semiconductors.

In this Letter, we develop such a model by coupling the NC measurements of the annealing
of low-energy, low-fluence implantation-induced disorder in \emph{c}-Si with results from
simulations performed using the recently proposed kinetic Activation-Relaxation Technique.
K-ART is an off-lattice, self-learning kinetic Monte Carlo
algorithm~\cite{ElMellouhi2008,Brommer2011} that makes it possible to follow the time
evolution of large implanted simulation boxes over a timescale of up to a second or more,
taking into account the full complexity of activated events and their associated
long-range elastic effects over timescales comparable the experiments. Results show that
the broad, featureless heat release as a function of temperature measured by NC over
several hundreds of Kelvin is well reproduced by the k-ART simulations. Analysis of the
atomistic simulations allows us to understand the microscopic origin of this logarithmic
relaxation.

NC measures heat release or absorption as a function of temperature in thermodynamic and
kinetic processes occurring at the nanoscale. The technique, described in
Refs.~\cite{1aN15,1aN10}, has been used to investigate phenomena ranging from melting
point depression in nanoparticles \cite{1aN3} to disorder annealing in amorphous Si
(\emph{a}-Si) \cite{1aN9} and polycrystalline Si (poly-Si) \cite{1aN8}. The device
consists of a low-stress, 250 nm thick SiN$_x$ membrane supported by a 300 $\mu$m thick
silicon frame. On top of the membrane, a Pt strip is used to measure the temperature and
heat the sample. The new device used here \cite{2aN15} features, underneath the membrane,
a 330 nm thick \emph{c}-Si strip, aligned with the heating strip. Heating rates reaching
10$^5$ K/s convert small amounts of heat released during disorder annealing into
measurable power.

For these experiments, the sample and reference NCs were in contact with a sample holder
either maintained at room temperature (RT) or cooled with liquid nitrogen (LN). In the
latter case, the membrane, surrounded by a thermal shield, reached 110 K. The silicon
layer was implanted with Si$^{-}$ at a fluence of 0.02--0.1 Si/nm$^{2}$ and an energy
ranging from 10 to 100 keV. The current was measured and integrated with systematic error
$<$20\%, although the relative uncertainty between different experiments within a series
of implantations is much smaller. A slit was placed in front of the NC in order to implant
the \emph{c}-Si strip without damaging the SiN$_x$ membrane on each side. The implanted
area being 0.05$\times$0.55 cm$^{2}$, 55 to 270 billion ions were implanted in each
experiment. This corresponds to 0.1 ions per square nanometer. At such low fluence, there
should be a very small proportion of collision cascade overlapping, so the experiments are
comparable to simulations of single ion implantations.
A delay of the order of 30 s occurred between the end of the implantation and the NC
temperature scan, which lasts 10 ms. Heat release is measured during the first temperature
scan. Subsequent scans are used as a baseline experiments to account for the fact that the
implanted and reference NCs are not identical.\cite{1aN15,1aN10}

\begin{figure} [htb!]
%\subfigure[]{
%   \includegraphics[width=.45\textwidth]{Si_recuit_12.pdf}
 %  \label{fig:nanocal_a}
 %  }
%\subfigure[]{
%   \includegraphics[width=.45\textwidth]{Si_recuit_13.pdf}
%   \label{fig:nanocal_b}
%   }
\includegraphics[width=.45\textwidth]{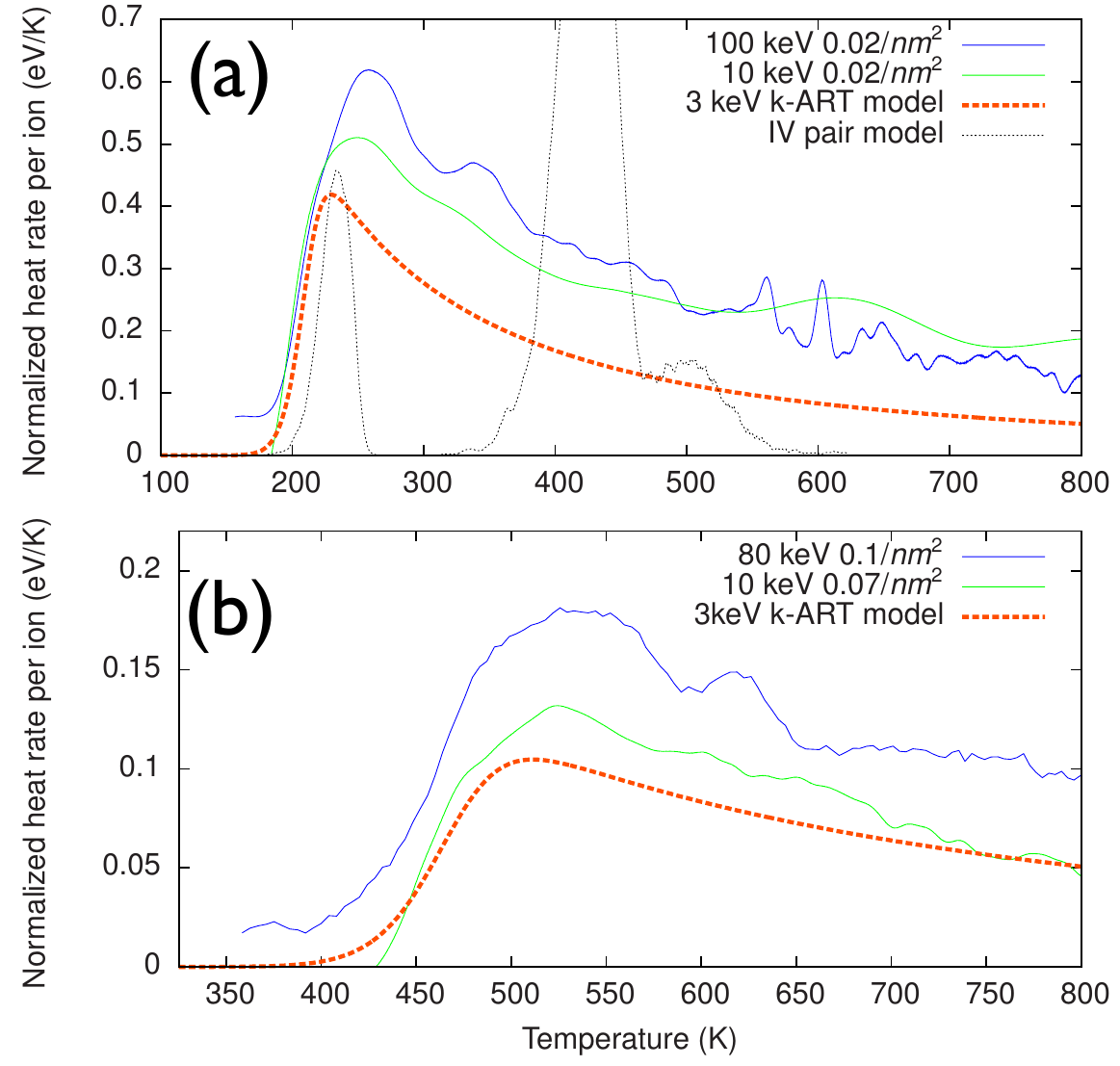}
%   \label{fig:figure_0}
  \caption{Heat release as a function of temperature starting from (a)
  liquid nitrogen (LN) and (b) at room temperature. Solid lines:
  Experimenta results; the NC signals are divided by the number of implanted ions and
  scaled to 3 keV (see text). Dashed red lines: model derived from k-ART after 3-keV Si
  ion implantation. Dotted black curve: simulation based on the IV-pair model at LN
  temperature.}

  \label{fig:nanocal}
\end{figure}

Figure \ref{fig:nanocal} shows as solid lines the heat release per unit temperature,
$dQ/dT$, as a function of temperature, for different implantations. The amount of heat
released is divided by the number of implanted ions. In order to compare with simulations
below, results are scaled to 3 keV, i.e., they are multiplied by a ratio of 3 keV divided
by the implantation energy. The data collected for 10 keV, 0.02 Si/nm$^{2}$ implants
featured a raw amplitude of less than 7 nJ/K above 500 K, a region where the signal
becomes dominated by the thermal losses, therefore featuring significant noise, so a
smoothed curve is presented. For implantations at LN, there is a rapid increase near 200
K, followed by a slow, featureless decrease spanning several hundred degrees. Heat
releases at similar amplitudes are observed for RT implants, but starting above 400 K. At
higher fluences, the signal becomes flat (not shown) instead of slowly
decreasing~\cite{anahory2011mecanismes}, a behaviour observed in
poly-Si \cite{1aN8} and
\emph{a}-Si \cite{1aN9}.

These measurements suggest that the disorder consists of structures complex enough that no
single activation energy can be associated to their annealing. Indeed similar experiments
involving H implantation, where discrete processes occur,
do show distinct peaks, associated to vacancy mobility, on top of a broad background
signal, related to more complex processes~\cite{anahory2011mecanismes}.

We first compare experiments to the IV-pair model proposed by Marqu\'{e}s \cite{Marques03}
using as input the list of defects predicted by SRIM with an 8w keV
implantation~\cite{SRIM}. Defects are then evolved using lattice-based kinetic MC
annealing during and after implantation with an activation energy $E_{\rm act}$ that
depends on the number $n$ of nearest neighbours in the ``IV-pair'' state \cite{Marques03}.
The simulation involves a 30~s waiting period at the lowest temperature to account for
the delay between the end of the implantation and the beginning of the NC temperature
scan. A typical result of our implementation of this model, with a displacement energy of
5 eV and at liquid nitrogen temperature, is presented in Fig.\ \ref{fig:nanocal} (a) (dotted
curve) with the signal scaled by 3/80 to compare to experiments and other models. The
simulation shows an isolated peak at 230 K, associated with the annealing of isolated IV
pairs ($n=0$) associated with activation energy of 0.43 eV, followed by a series of peaks
corresponding to the other activation energies ($n>0$), and fails to reproduce
experiments. Clearly, important relaxation mechanisms are missing from this lattice-based
KMC model.

To attempt to better capture the complex relaxation between the various structures forming
the disorder, we turn to the kinetic Activation-Relaxation Technique,
k-ART, an off-lattice self-learning kinetic Monte Carlo method. K-ART couples ART nouveau
for the event search with a topological analysis tool for the catalog building and with a
KMC algorithm for the time propagation~\cite{ElMellouhi2008,Brommer2011}. By performing an
extensive search for saddle points and fully relaxing the relevant energy barriers before
each event, k-ART is able to take into account short and long-range elastic effects. The
topological analysis, performed with NAUTY \cite{McKay:1981:45}, allows a stable and
reliable management of events even for disordered and complex configurations such as
vacancy diffusion in Fe \cite{PhysRevLett.108.219601} and relaxation in amorphous
silicon~\cite{Brommer2011}.

We first inject a 3~keV Si atom into a 300~K 100 000-atom slab of Stillinger-Weber silicon
with a Langevin bath as boundary condition perpendicular to the trajectory of the
implanted atom, simulating the effect of a low-energy low-fluence ion implantation
experiment. The cell is then relaxed using MD for one to ten ns. A 27 000-atom subpart of
the cell containing the defects is then cut out and placed into a box with periodic
boundary conditions along all direction, to eliminate surface effects. We generated three
independent samples following this procedure and then launched several 300~K k-ART
simulations on each MD-produced cascade, simulating up to timescales of 1~s or more.

\begin{figure} %[htb!]
\centering
%  \includegraphics[width=0.4 \textwidth]{finalConf2.pdf}\\
 % \caption{{Typical configuration of a system after 1 s. An interstitial (I) and a vacancy (V) are identified. }}
%\subfigure[]{
 %  \includegraphics[width=.35\textwidth]{Si_recuit.pdf}
 %  \label{fig:kart_a}
 %  }
%\subfigure[]{
%   \includegraphics[width=.35\textwidth]{mon_graph_2.pdf}
%   \label{fig:kart_b}
%   }
\includegraphics[width=.40\textwidth]{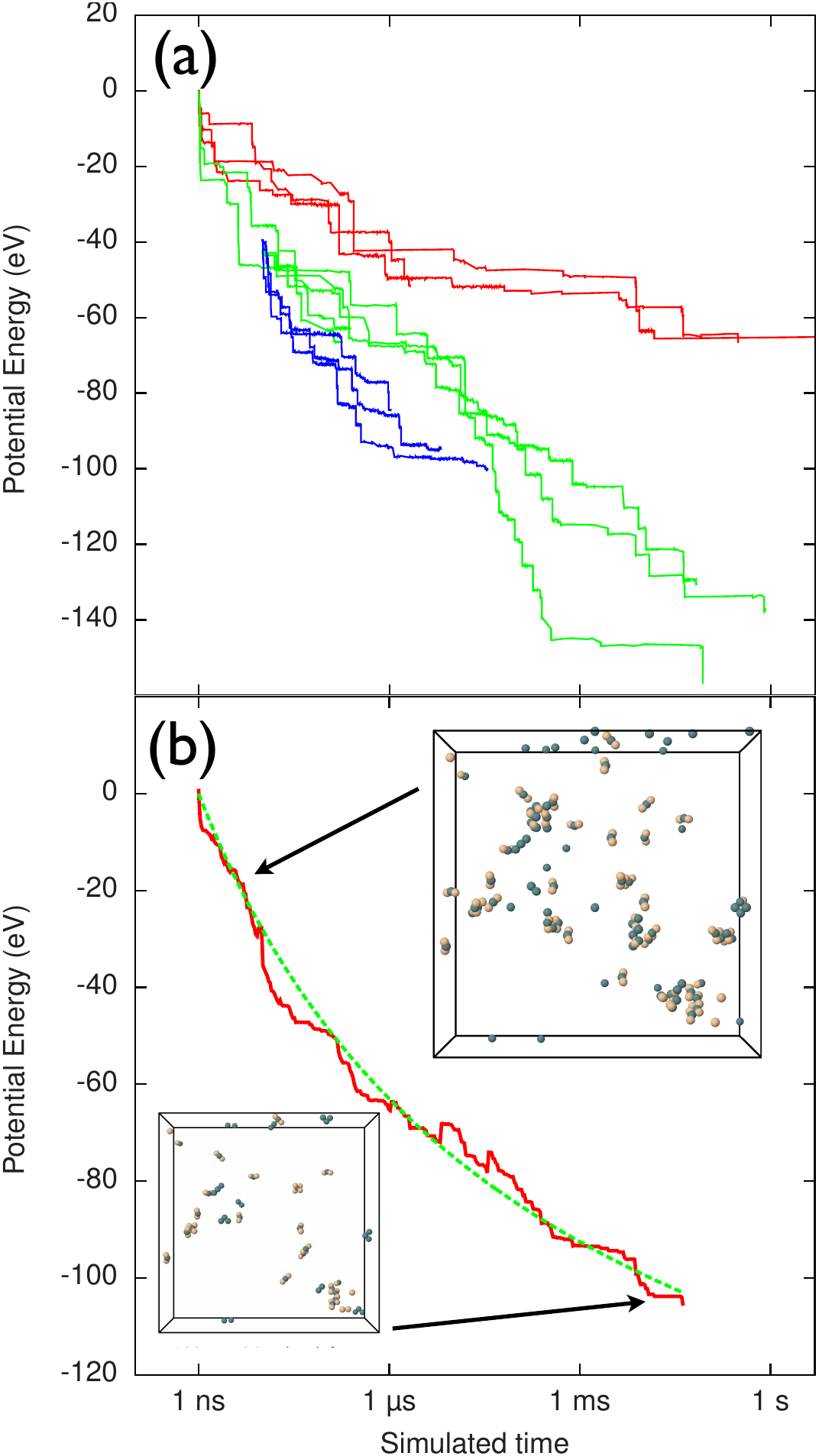}

\caption{(a)Energy evolution for k-ART simulations. The lines correspond to independent simulations starting from three different implentation runs. (b) Red solid line: Average of the curves in panel (a). Green dashed curve: Solution of Eq.~\ref{eq:basic_derivatives} at $T=300$~K. Insets: typical configurations of defects in a 27000-atoms system after a 10 ns MD run (upper-right panel; there are184 point defects) and after a 0.1 s k-ART simulation(lower-left panel; there are 94 point defects). Interstitials are colored in beige and vacancies in blue.
}

\label{fig:kart}
\end{figure}

Energy evolution for the k-ART trajectories is shown in Fig.~\ref{fig:kart} (a). Each
relaxation simulation shows a logarithmic time dependance over many orders of magnitude
with a slope determined by the initial disordered configuration; independent k-ART
simulations starting from the same initial model follow a similar relaxation pathway. Such
logarithmic relation is similar to that observed in heat release and relaxation experiment
in other complex systems such as polymer glasses~\cite{Knoll2009}. After averaging over
the various simulations, in order to reproduce experimental measures over a large number
of cascades, we find a slightly curved but still logarithmic overall relaxation behavior
for the system that is robust to the addition or subtraction of other runs
(Fig.~\ref{fig:kart} (b)).

Before examining in more details the atomistic mechanisms leading to the macroscopic
relaxation, we need to confirm that the simulation results reproduce the experimental
data. For this, it is necessary to convert our fixed temperature results into
constant-heating energy release. We consider an initial density of processes $n(E,t=0)$
that can be activated by first order kinetics. The number of activated events with a
barrier between $E$ and $E + dE$ during the time interval between $t$ and $t+dt$ is
\begin{equation}
        \label{eq:basic_derivatives}
	dn(E,t) dE dt = -n(E,t) \nu e^{-E/kT} dE dt.
\end{equation}
This equation can be solved at fixed (similar to the simulation) or increasing temperature
(such as in NC scans).

Following the analysis of the accepted events, and in agreement with previous
work~\cite{Kallel10}, we assume that activated processes release a fixed amount of energy
$h_0$, independent of barrier height. A direct analysis of the microscopic events
indicates that the effective $n(E,0)$, which is kinetically determined, goes like
$n(E,0)\propto E^{\mu}$ with $-2.0< \mu<-1.2$. To improve accuracy, we fit
Eq.~\ref{eq:basic_derivatives} to the averaged fixed temperature energy relaxation shown
in Fig.~\ref{fig:kart} (b) and find that the event density per interval $log(t)$ decreases
with increasing time, in agreement with the direct estimation, and that
$h_0n(t=0)=53\,E^{-1.7}$. With this distribution, we compute the heat released under
experimental conditions, including an initial annealing period of 30 s at the implantation
temperature to reflect experimental conditions. Results are shown in
Figs.~\ref{fig:nanocal} (a) and \ref{fig:nanocal} (b). The agreement between
simulation-derived energy release and experiments both at liquid nitrogen (LN) and room
temperature (RT) is excellent, especially considering that the simulations are performed
with an empirical potential.

The correspondence between simulations and experiments is non-trivial. A uniform density
of process, $n(E,t=0)$, for example would lead to a flat heat release such as that
observed in higher implantation fluences~\cite{anahory2011mecanismes} and
\emph{a}-Si~\cite{1aN8,Kallel10}. The origin of the exponent in the density of processes
is therefore associated with a decrease in the number of available barriers as defects
anneal, a behaviour that does not take place when amorphous pockets remain present. This
decrease in complexity is observed in the microscopic evolution. After 1 ns following the
implantation, before the k-ART simulation, the various models contain between 100 and 150
point defects (I or V) assembled into 20 to 30 clusters. Although these clusters vary in
size from 1 to 30 point defects, most of them contain between 2 and 5 defects and are
therefore better classified as defect complexes rather than amorphous pockets. After 1~ms
to 1~s, the various simulations lead to a range of configurations that consist, typically,
of small defect complexes, with 3 to 5 point defects, and only a few clusters. This is
illustrated in the insets of Fig.~\ref{fig:kart} (b), that shows typical configurations
after 10 ns and 0.1 s.

% We identify three main classes of energy relaxation mechanisms: recombination events
%(annihilation of point defects), accumulation events (aggregation of point defects) and
% reconfiguration events (clusters of point defects that change configuration). Accumulation
% of defects and reconfiguration of clusters generally allow small drops in potential energy
% (in the order of one eV or less) as well as pinning otherwise mobile defect structures.
% Recombination events correspond to large drops in energy, of the order of several eV, some
% of them corresponding to an IV-pair (or bond defect) recombination. In the absence of
% other nearby defects, we report an energy barrier to bond defect recombination in
% agreement with the literature (0.43 eV). Accumulation of these bond defects does not
% necessarily lead to the higher recombination barrier postulated by nucleation
% theory~\cite{Marques03}, it depends rather on the precise orientation of the defect
% cluster at hand. The effect can increase, decrease, or have no impact on the stability of
% the bond defect. While all these moves are essential for relaxing the disordered region,
% their direct contribution to energy release is unequal. Overall, recombination events
% account for 40~\% of events and 85~\% of heat emission, while accumulation and
% reconfiguration events account for 60~\% of events and 15~\% of the heat emission.
% 

\begin{figure} %[htb!]
\centering
%\subfigure[]{
 %  \includegraphics[width=.38\textwidth]{Si_recuit_2.pdf}
 %  \label{fig:events_a}
 %  }
%\subfigure[]{
 %  \includegraphics[width=.38\textwidth]{Si_recuit_7.pdf}
 %  \label{fig:events_b}
 %  }

\includegraphics[width=.40\textwidth]{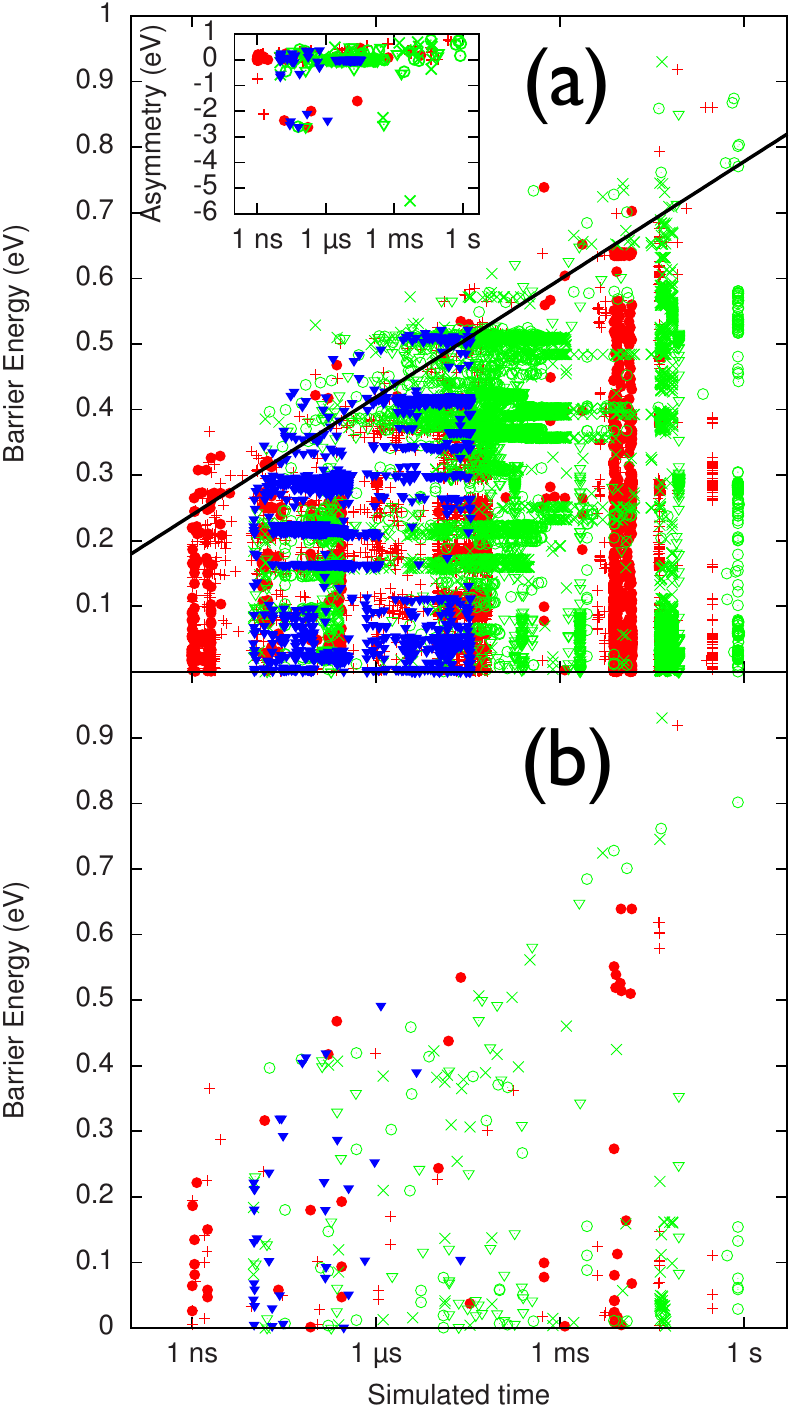}
  \caption{(a) Energy barrier of all executed k-ART events. Points above the black line correspond to the 10 \% highest energy barriers in each time-frame. Inset: the potential energy asymmetry (final energy minus initial energy) of the executed events with an energy barrier lying in the top 10 \% of the barriers plotted in Fig.~\ref{fig:events} (a) for each time-frame. (b) Energy barrier of the k-ART events that release at least 0.5 eV of heat.}
  \label{fig:events}
\end{figure}

Structural information is not sufficient to explain the logarithmic relaxation.
Figure~\ref{fig:events}(a) shows the energy barrier separating each state selected by
k-ART, aggregated over the simulations that reached at least 10 $\mu s$, as a function of
time. Two features stand out. First, a quasi-continuous distribution of activated events
is accessible in all time frames. Even after relaxing the system by several tens of eV,
low-barrier events are still present and executed. Second, the maximum energy barrier for
executed events increases logarithmically with time.

The quasi-continuous barrier distribution suggests that the system can be
kinetically-limited by configurational entropy. Most of the low-energy barriers connect
flickering states that do not lead to structural evolution. The logarithmically growing
maximum energy barriers, for their part, indicate that the structural evolution is also
energy-limited and it is this interplay that generates a logarithmic energy decay.

The link between energy relaxation and kinetics is given by the high-energy barriers. The
inset of Fig. \ref{fig:events} (a) shows the heat released by the events with an energy
barrier above the 90$^{\text{th}}$ percentile, i.e.~those above the straight line. These
do not generally lead to large drops in potential energy and many even lead to a
higher-energy state, in agreement with recent observations that the forward and reverse
energy barriers are totally uncorrelated~\cite{Kallel10}; the energy barrier of events
that lead directly to a significant relaxation are distributed evenly throughout all the
selected events (Fig. \ref{fig:events} (b)). Our simulations indicate that the annealing
of the implanted \emph{c-}Si is not systematically kinetically limited by the relaxation
events, but limited by events that allow the system to leave a region of the
configurational space where the potential energy landscape is essentially flat to reach
another region where a large drop in potential energy is accessible. It can be described
as a two-step replenish and relax process that explains the explains the logarithmic
relaxation. High energy barrier events, which are mostly reconfiguration events, do not
directly lead to a low-energy structure but rather unlock the system, open new low-energy
pathways and replenish the basin of available energy-releasing events. These events are
associated with an almost continuous and time-independent energy barrier distribution that
reflects the complexity of the defects themselves and the impact of long-range elastic
deformations.

%Using both nanocalorimetry and second-long atomistic simulations to characterize the
%energy relaxation of ion-damaged c-Si, we show that the microscopic mechanisms
%associated with logarithmic relaxation take place as a two-step replenish and relax
%process associated with different time scales. As the system relaxes, it reaches states that are more metastable and logarithmically increasing barriers need to be overcome to find events that will lead to lower energy configurations.  

\begin{acknowledgments}
The authors are grateful to S. Roorda for fruitful discussion, to L. Godbout,
X. Perraton and R. Gosselin for their excellent technical assistance, M.
Skvarla and P. Infante of the Cornell Nanofabrication Facility, as well as M.
H. Bernier and P. Vasseur of \'{E}cole Polytechnique de Montr\'{e}al for
their assistance with NC fabrication. We thank Calcul Qu\'{e}bec for generous
allocation of computer ressources. This work benefited from the financial
support of NanoQu\'{e}bec, the Fonds qu\'{e}b\'{e}cois de recherche sur la
nature et les technologies, the Natural Science and Engineering Research
council of Canada and the Canada Research Chair Foundation. The K-ART software package can be obtained via NM.
\end{acknowledgments}

\bibliography{references}

\end{document}